\documentstyle[aps,preprint]{revtex}

\begin{document}
\author{V.Yu. Irkhin$^{*}$ and Yu.P. Irkhin}
\address{Institute of Metal Physics, 620219 Ekaterinburg, Russia}
\title{Charge screening and magnetic anisotropy in RCo$_5$ compounds}
\maketitle

\begin{abstract}
An analysis of magnetic anisotropy in RCo$_5$ compounds is performed with
account of screening of ion point charges by conduction electrons. A crucial
role of non-uniform distribution of screening electrons (the terms
containing derivatives of charge density) is demonstrated. Influence of
anisotropy of screening, that is connected with the anisotropy of the Fermi
surface, on the magnetic anisotropy sign is discussed.
\end{abstract}


It is believed now that the main mechanism of occurrence of strong magnetic
anisotropy (MA) in the rare-earth (RE) based intermetallic systems is the
crystal field one\cite{irk,bu}. The simplest point-charge model leads
frequently to difficulties and contradictions with experimental data. For
example, the calculated MA constant $K_1$ in RCo$_5$ compounds turns out to
be very large and have an incorrect sign\cite{ros}. Similar difficulties
occur for the R$_2$Fe$_{14}$B systems\cite{adam}. Thus screening of the
crystal field should play an important role. This screening can be taken
into account by introducing the effective ion charge $Q^{*}$ which can
differ considerably from the bare ion charge. Shielding of crystal fields in
ionic solids\cite{stern} turns out to be insufficient even for qualitative
explaining experimental data.

Another approach to the MA problem is based on the first-principle
band-structure calculations. Generally, RE systems are a difficult case for
standard band theory methods, so that one has to use some approximations and
model representations (e.g., atomic sphere approximation in early papers,
inclusion of $f$-states in the core states). Modern calculations\cite
{richt,fahn} treat mainly the anisotropic contribution of the conduction
electrons in the atomic sphere with the centre at the RE $f$-ion, and the
rest ``lattice'' contribution is obtained in the point-charge approximation
with screened charge densities of other cells, which do not overlap with the 
$f$-shell. It should be noted that the atomic and lattice contribution are
in fact not independent, and their partial cancellation takes place. The
local-spin-density full-potential approaches, which are based on total
energy calculations\cite{fahn}, can provide correct orders of magnitude of
anisotropy energy, but quantitative agreement with experimental data is not
satisfactory.

Thus simple qualitative considerations which start from the physically
transparent point-charge picture, but introduce effective distance-dependent
ion charges, seem still be useful. In the present paper we analyze the first
MA constant with account of charge screening by conduction electrons in a
metal. As demonstrated in Ref.\cite{ii}, effective charges depend strongly
on the concrete form of screening electron density. In particular, the
Friedel oscillations of charge density can play an important role. Here we
discuss the screening in more details, in particular with account of its
anisotropic character.

The MA constants are determined from angle dependence of the energy of
magnetic ions in the crystal field 
\begin{equation}
\delta {\cal E}_{cf}=-K_1\cos ^2\theta +...  \label{an}
\end{equation}
We consider the magnetic ion at the point ${\bf r}=0$ in the crystal field
of the surrounding charges. The contribution to the crystal-field potential
from the ion with the bare charge $Q_0$ at the point {\bf R }can be
represented as 
\begin{equation}
V_{cf}({\bf r})=\frac{Q_0+Q_{el}({\bf R-r})}{|{\bf R-r}|}  \label{V}
\end{equation}
where $Q_{el}({\bf R})$ is the conduction-electron screening charge. After
expanding (\ref{V}) in $r$ up to $r^2$, the expression for $K_1$ can be
written in the form (cf. Refs.\cite{ros,ii}) 
\begin{equation}
K_1^{pc}=3e^2\Lambda \langle r_f^2\rangle \alpha _JJ(J-1/2)  \label{pc}
\end{equation}
Here $\langle r_f^2\rangle $ is the average square of the $f$-shell radius, $%
J$ is the total angular momentum of the RE ion, $\alpha _J$ is the Stevens
factor, 
\begin{equation}
\Lambda =\sum_{{\bf R}}Q^{*}({\bf R})\frac{3\cos ^2\theta _{{\bf R}}-1}{R^3},
\end{equation}
the sum goes over the lattice, $\theta _{{\bf R}}$ is the polar angle of the
vector ${\bf R}$, $Q^{*}({\bf R})$ are the corresponding screened ion
charges. Note the difference of our consideration from Ref.\cite{richt},
where the screening charge in the sphere with the centre at the magnetic RE
ion was calculated.

In the spherical charge density approximation we have\cite{ii} 
\begin{equation}
Q^{*}({\bf R})=Q_0+Q_{el}(R)-\frac 43\pi R^3[Z(R)-RZ^{\prime }(R)]
\label{qef}
\end{equation}
where $Q_{el}(R)$ is the conduction electron charge inside the sphere with
the centre at the point charge and radius $R$, 
\begin{equation}
Q_{el}(R)=4\pi \int_0^R\rho ^2d\rho Z(\rho )  \label{qq}
\end{equation}
$Z(R)$ is the charge density, $Q_{el}^{\prime }(R)=4\pi r^2Z(R)$, the system
of units where the electron charge $e=-1$ being used. We see that $Q^{*}(%
{\bf R})$ includes explicitly, besides the total charge $Q_{el}(R),$ also
the charge density $Z(R)$ and its derivative $Z^{\prime }(R).$ Such terms do
not occur in the calculations\cite{richt} where the ``lattice'' charge
density does not overlap with the $f$-shell. Note that higher-order
anisotropy constants are obtained after keeping next-order terms in $r$ and
include higher-order derivatives of $Z(R).$

To obtain the value of $Q^{*}({\bf R}),$ one has to investigate the charge
screening for a concrete electronic spectrum. In Ref.\cite{ii}, the
one-centre screening problem was considered within a simple model of free
conduction electrons ($E=k^2/2$) in the rectangular potential well which is
induced by single impurity and has the width $\ d$ and depth $E_0=k_0^2/2$%
\cite{dan}. This model enables one to calculate the charge distribution of
screening conduction electrons in terms of the spherical Bessel and Neumann
functions $j_l(kr)$ and $n_l(kr)$ ($r>d$) and the scattering phase shifts $%
\eta _l$. The value of $k_0$ is determined for given $k_F$ and$\ d$ from the
Friedel sum rule 
\begin{equation}
Q_0=\frac 2\pi \sum_{l=0}^\infty (2l+1)\eta _l(k_F)  \label{fri}
\end{equation}
The parameter $d$ should be determined by the geometry of the lattice near
the impurity. In Ref.\cite{dan}, where impurities in the Ag host were
considered, $d$ was chosen to be equal to the Wigner-Seitz radius, so that $%
k_Fd=2$. The results of the calculations for $k_Fd=2$ and $k_Fd=3$ are
presented in Figs.1,2. At $Q_0=1$ Eq.(\ref{fri}) yields $k_0d=1.46$ and $%
k_0d=1.235$ respectively.

The model discussed is more adequate for impurities which induce strong
disturbance of charge density (e.g., in hydrogen-containing RE systems\cite
{nik}). Of course, the choice of $d$ may be different and more complicated
models are required for a regular lattice of screened charges where
interference of screening charge clouds from different centres takes place.
In metals, the value of ion charge $Q_0$ can be put to the charge in some
sphere (e.g., charge transfers in the atomic sphere were considered in Ref.%
\cite{richt}) and does not necessarily coincide with the nominal free ion
value. Besides that, the dependence $Q^{*}({\bf R})$ in the lattice can
become anisotropic. Nevertheless, we use Figs.1-2 for a qualitative
discussion.

One can see from Fig.2 that at $R<d$, except for the case of very small $R$
where $Q^{*}(R)$ slightly decreases, the derivative term in (\ref{qef})
results in that $Q^{*}(R)$ grows (despite an increase of $|Q_{el}(R)|$). For 
$R\simeq \ d,$ where $Z^{\prime }$ is maximum, the non-uniform distribution
of electron density leads to that the effective ion charge $Q^{*}$ is
positive and exceeds considerably its bare value $Q_0$ ($Q_0=1$ in our
case). At the same time, with further increasing $R$ the situation changes
drastically: $Z^{\prime }$ decreases and becomes negative, so that
``overscreening'' of the ion charge occurs. At large distances $Q^{*}$ tends
to zero, but considerable oscillations of the effective charge sign take
place, which attenuate rather slowly. It should be stressed that the
oscillations are due mainly to the derivative term. On the other hand, the
quantity $Q_0+Q_{el}(R)$ monotonically decreases up to $R\simeq d$ and then
tends to zero very rapidly, oscillations being very weak (Fig.1). This
agrees with the fact that band calculations yield usually small values of
charge transfers in atomic spheres (see, e.g.,\cite{richt}). The oscillation
period and position of maxima and minima of $Q^{*}(R)$ turn out to be weakly
sensitive to the values of $Q_0$ and $d$, but are determined mainly by the
parameter $k_F.$

Now we analyze concrete geometry of the RCo$_5$ lattice (CaCu$_5$ structure,
Fig.3) with $\,c\simeq 4$\AA $,a\simeq 5$\AA . The Co ions have two
positions in two different types of hexagonal layers: 2c (CoI) sites in the
layers containing R atoms with $R=a/\sqrt{3}\simeq 0.57a$ and 3g (CoII)
sites in layers with no R atoms with $R=\frac 12\sqrt{a^2+c^2}\simeq 0.64a.$
Although positions of all R ions are equivalent, the contributions to the
crystal field at a given R site from the charges in the same plane (sites
RI, $R=a$) and in neighbor planes (sites RII, $R=c$) can be different, since
the effective charge is a function of the distance between two sites. Then
we can write down (cf.\cite{ros}) 
\begin{equation}
\Lambda =6a^{-3}\left( 16\frac{2y^2-1}{(1+y^2)^{5/2}}Q_{\text{CoII}%
}^{*}-3^{3/2}Q_{\text{CoI}}^{*}+\frac 23y^{-3}Q_{\text{RII}}^{*}-Q_{\text{RI}%
}^{*}\right)  \label{co}
\end{equation}
where $y=c/a\simeq 0.8.$ The contributions of in-plane (I) and out-of-plane
(II) neighbors of R site have different signs. However, unlike the pure rare
earth-metals with the hcp structure\cite{irk}, the small geometrical factor $%
\sqrt{8/3}-c/a=1.633-c/a\sim 0.05$ does not occur, so that the calculated
value of $K_1$ turns out two orders of magnitude larger.

The experimental data on the $c/a$ ration and first magnetic anisotropy
constant of the RCo$_{5+x}$ compounds at low temperatures are presented in
Table 1. A monotonic decrease of $c/a$ in the RE series takes place (with
exception of tetravalent Ce).

Table 1. The total angular momenta $J$, Stevens factors $\alpha _J,$ average
squares of the $f$-shell radius $\langle r_f^2\rangle $ (atomic units) for
free R ions; $c/a$ ratios and the experimental values of $K_1$ (K/RE ion) in
RCo$_{5+x}$ systems [for light R according to\cite{erm1,sank} and for heavy
R (the contribution of RE sublattice) according to Ref.\cite{Erm}]. The
corresponding values of $\overline{Q}_{\text{Co}}^{*}$ are calculated by
using (\ref{l23}), (\ref{pc}).

$
\begin{tabular}{|lllllllll|}
\hline
\multicolumn{1}{|l|}{RCo$_{5+x}$} & CeCo$_5$ & PrCo$_5$ & NdCo$_5$ & SmCo$_5$
& \multicolumn{1}{l|}{TbCo$_{5.1}$} & \multicolumn{1}{l|}{DyCo$_{5.2}$} & 
\multicolumn{1}{l|}{HoCo$_{5.2}$} & ErCo$_6$ \\ \hline
\multicolumn{1}{|l|}{$J$} & 5/2 & 4 & 9/2 & 5/2 & \multicolumn{1}{l|}{6} & 
\multicolumn{1}{l|}{15/2} & \multicolumn{1}{l|}{8} & 15/2 \\ \hline
\multicolumn{1}{|l|}{$\alpha _J\times 100$} & -5.7 & -3.4 & -7.1 & 4.1 & 
\multicolumn{1}{l|}{-1.0} & \multicolumn{1}{l|}{-0.63} & \multicolumn{1}{l|}{
-0.22} & 0.25 \\ \hline
\multicolumn{1}{|l|}{$\langle r_f^2\rangle $} & 1.20 & 1.09 & 1.00 & 0.88 & 
\multicolumn{1}{l|}{0.76} & \multicolumn{1}{l|}{0.73} & \multicolumn{1}{l|}{
0.69} & 0.67 \\ \hline
$c/a$ & 0.817 & 0.797 & 0.796 & 0.795 & 0.803 & 0.810 & 0.817 & 0.821 \\ 
\hline
\multicolumn{1}{|l|}{$K_1^{\exp }$} & -61 & -44 & -220 & 190 & 
\multicolumn{1}{l|}{-96} & \multicolumn{1}{l|}{-211} & \multicolumn{1}{l|}{
-203} & 80 \\ \hline
\multicolumn{1}{|l|}{$\overline{Q}_{\text{Co}}^{*}$} & -0.007 & -0.003 & 
-0.006 & -0.04 & \multicolumn{1}{l|}{-0.02} & \multicolumn{1}{l|}{-0.03} & 
\multicolumn{1}{l|}{-0.09} & -0.04 \\ \hline
\end{tabular}
$

One can see that the experimental values $K_1^{\exp }$ for heavy rare earths
and Sm are obtained for $\Lambda a^3\simeq 1.$ Taking into account only the
contributions of Co ions in (\ref{co}) and assuming $Q_{\text{CoI}}^{*}=Q_{%
\text{CoII}}^{*}=\overline{Q}_{\text{Co}}^{*}$ we obtain 
\begin{equation}
\Lambda \simeq -\frac{23.4}{a^3}\overline{Q}_{\text{Co}}^{*}  \label{l23}
\end{equation}
Thus, for $\overline{Q}_{\text{Co}}^{*}\sim 1,$ $K_1$ has incorrect sign and
is very large in absolute value (of order of 3000K). It should be noted that
such values of $K_1$ are in fact not self-consistent: they would destroy the
Russel-Saunders coupling and quench total momenta. In light rare earths (Ce,
Pr, Nd and Sm) the momenta, as obtained from both neutron scattering and
magnetization measurements\cite{jap}, are indeed considerably suppressed. Ce
ions are supposed to be tetravalent; for other light rare earths, effects of
the strong crystal fields and exchange interactions can play a role. On the
other hand, in Tb, Dy, Ho and Er the saturation momenta are close to their
free ion values\cite{jap1}.

To compensate large numerical factor in (\ref{l23}), one has to put $%
\overline{Q}_{\text{Co}}^{*}$ to be very small in absolute value and
negative (Table 1). One cannot rule out that the distance between Co ions
corresponds to negative values of $Q^{*}$ with $|Q^{*}|\ll 1$ in Fig.2, but
such a situation is rather unusual. Thus the simple model with equal Co
charges can hardly explain the observed sign and value of $K_1,$ and a more
detailed treatment of screening is needed.

An assumption that main role in the crystal field formation belongs to R
ions and $Q_{\text{Co}}^{*}$ can be assumed zero was made in Refs.\cite
{greed,sank}. The effective charges for R ions, $Q_{\text{R}}^{*},$ could be
supposed to equal about 1, as well as in pure rare earths where this
parameter varies between 1.1 and 1.4\cite{ii}. Then we would obtain the
correct sign and order of magnitude for $K_1$ since the contribution of RII
ions dominates. However, the distances between RE ions in RCo$_5,$ $c$ and $%
a,$ are considerably larger than in pure rare earths (about 3.5\AA ), so
that the values $Q_{\text{R}}^{*}$ can be much smaller. Besides that, the
variation of effective charge in the RCo$_5$ series is much stronger than
for pure rare earths. Such a variation can be related to the contribution of
CoII sites since the factor at the first term in the brackets of (\ref{co})
depends appreciably on $y=c/a.$

It is instructive to take into account the anisotropy of screening of Co
charges that is connected with the anisotropy of the Fermi surface in the
hexagonal lattice. Then the screening is determined by the wavevector of the
Fermi surface in the corresponding direction. The effective values of $k_F$
in the hexagonal planes can be supposed smaller than in the direction to
CoII sites. Then the charge of the CoII ions is screened with distance
slower than for CoI ions and can dominate in $K_1$, despite the larger
distance and numerical factor in (\ref{co}). This circumstance may lead also
to additional (besides pure geometrical) strong dependence of the anisotropy
on $y.$ It should be noted that the calculation for SmCo$_5$\cite{richt}
yields for charge transfers in the Co atomic spheres (as well as for CoI and
CoII magnetic moments) different values: $q_{\text{CoI}}\simeq 0,$ $q_{\text{%
CoII}}\simeq -0.03.$

Experimental values of $K_1$ from Table 1 can be obtained, e.g., for $Q_{%
\text{CoII}}^{*}\simeq 4Q_{\text{CoI}}^{*}\simeq 0.3,$ $Q_{\text{R}}^{*}=0,$
or for $Q_{\text{CoII}}^{*}\simeq 3Q_{\text{CoI}}^{*}\simeq 0.2,$ $Q_{\text{R%
}}^{*}\simeq 1$ (in the latter case, the sign of $K_1$ is determined by R
contribution which is partially cancelled by the Co contribution). Thus
occurrence of very small values of $Q_{\text{Co}}^{*}$ by accidental reasons
is not required in such a consideration.

To conclude, an account of distance-dependent screening of ion charges by
conduction electrons can provide a reasonable explanation of the first MA
constant value in RCo$_5$ compounds, the non-uniform distribution of
screening electrons (the derivative term in the expression for the effective
charge (\ref{qef})) being of crucial importance. Our approach does not
introduce artificial separation into atomic spheres. On the other hand, it
has a number of other drawbacks. In particular, it treats independent charge
centres and is not self-consistent: perturbation of charge density by R ion
itself (see, e.g., Ref.\cite{ros}) is not taken into account. Thus a
synthesis of our consideration with approaches, that are based on the real
electronic structure calculations, would be of interest for further insight
into the problem of magnetic anisotropy in the rare-earth systems.

{\sc Figure captions}

Fig.1. The distance dependence of the sum of the bare ion charge and
electron screening charge, $Q_0+Q_{el}(R),$ for $k_Fd=2$ (solid line) and $%
k_Fd=3$ (dashed line) with $Q_0=1.$

Fig.2. The distance dependence of the effective charge $Q^{*}(R)$ (\ref{qef}%
) for the same parameter values as in Fig.1.

Fig.3. The local environment of the R site in the crystal structure of RCo$%
_5 $ compounds. Large open circles denote R sites, small open circles CoI
(2c) sites, small shaded circles CoII (3g) sites.

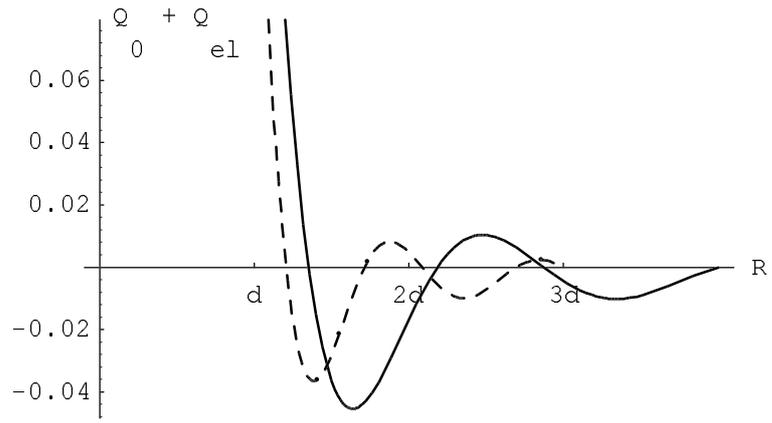
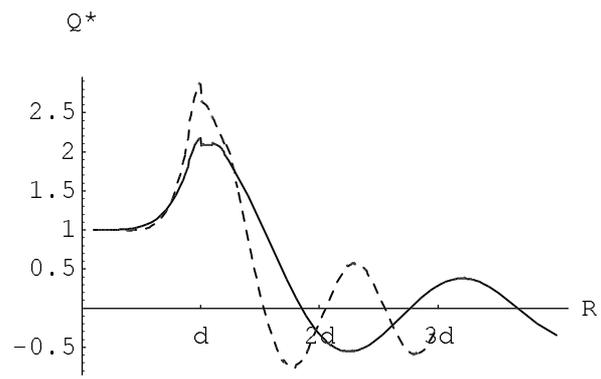
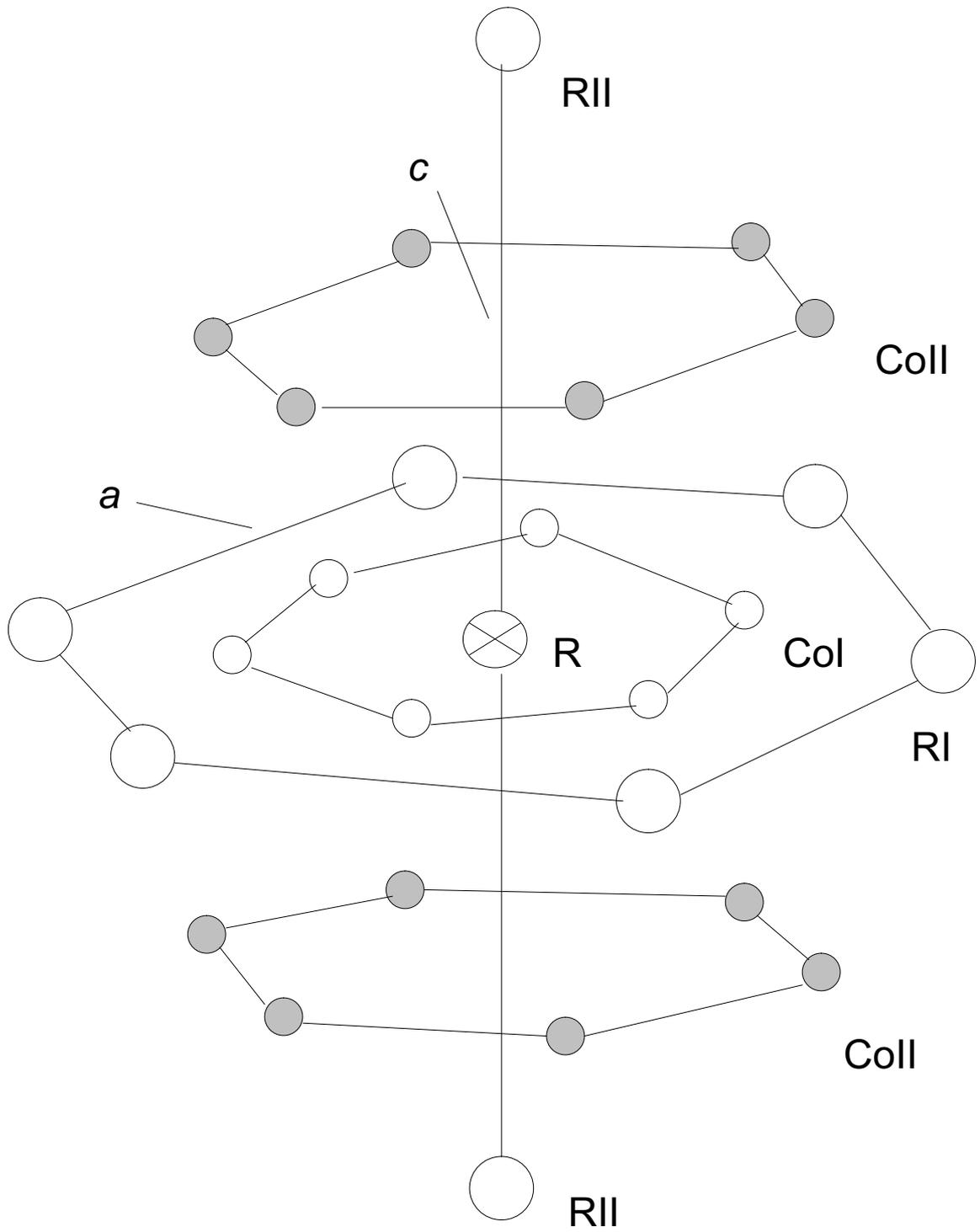
\end{document}